\documentclass[a4paper,11pt]{article}
\usepackage{pos}
\usepackage{lineno}
\usepackage{enumitem}
\usepackage{wrapfig}
\title{Expanding the Pierre Auger Observatory Open Data program}
\author*[a]{Viviana Scherini}
\affiliation[a]{Universit\`a del Salento and INFN, Lecce, Italy}

\onbehalf{for the Pierre Auger Collaboration$^b$}
\affiliation[b]{Observatorio Pierre Auger, Av.\ San Mart{\'\i}n Norte 304, 5613 Malarg\"ue, Argentina\\
Full author list: {\rm\url{https://www.auger.org/archive/authors_icrc_2025.html}}}

\emailAdd{spokespersons@auger.org}

\abstract{Since 2021, the Open Data Portal has provided access to the Pierre Auger Observatory’s data for
both the scientific community and the general public. The data release process has been in place
since the Observatory's foundation. It continues to be strengthened as outlined in the approved
policy and the Observatory’s Data Management Plan. More than 80\,000 cosmic-ray events above
$10^{17}$ eV, detected with the surface and fluorescence detectors, have been released at various
levels, from calibrated traces to high-level reconstruction parameters. Additionally, atmospheric
data and low-energy particle counting rates have been made available for space weather studies.

The Collaboration is committed to releasing FAIR (Findable, Accessible, Interoperable, and
Reusable) data, along with accompanying software and detailed documentation, enabling users to
perform their own queries and analyses for both research and educational purposes. These datasets
have already served as a basis for several scientific papers and have been widely used in various
outreach activities.

After 20 years of stable data acquisition, the Pierre Auger Collaboration will disclose 30\% of the
cosmic ray events above $2.5 \cdot 10^{18}$ eV collected with the main surface detector array between 2004 and 2022, corresponding to an exposure of about $24\,000$ km$^2$ sr yr, together
with events detected with the fluorescence detector and used for energy calibration. This release
will provide an unprecedented public dataset for ultra-high-energy cosmic rays, enabling in-depth
studies of their properties.

Together with the published catalog of the 100 most energetic events recorded, this initiative
represents the Pierre Auger Collaboration’s strong commitment to distributed and collective
knowledge, sharing progress with the entire scientific community.}

\ConferenceLogo{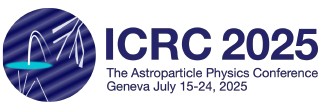}

\FullConference{39th International Cosmic Ray Conference (ICRC2025)\\
 15–24 July 2025\\
Geneva, Switzerland\\}
\begin{document}
\maketitle
%\linenumbers
\section{Introduction} 

After more than 20 years of data taking, the Pierre Auger Observatory~\cite{paoNIM}, the largest facility for the measurement of ultra-high energy cosmic rays (UHECR), has detected more than 20\,000 cosmic-ray events per year with an energy above 2.5~EeV (1 EeV = $10^{18}$ eV) providing, with unprecedented statistics and precision, major breakthroughs in the field.

%The Observatory is located on a high-altitude plain near Malarg\"ue, Mendoza Province, Argentina, at a mean altitude of about 1400 m, corresponding to an atmospheric overburden of about 875 g\,cm$^{-2}\,$.  It combines two detection techniques for the detection of extensive air showers (EAS). A Surface Detector array of 1600 water-Cherenkov detectors (WCD) in a 1500 m triangular grid, SD-1500, yielding an effective area of 3000 km$^2$, provides lateral sampling of the EAS at the ground. A Fluorescence Detector (FD) comprising 24 telescopes grouped at four sites (eyes) overlooks the array and detects the UV fluorescence light emitted by the de-excitation of the nitrogen molecules previously excited by the charged particles from the EAS. 

The rich data collected by the Collaboration covers different and complementary research fields from astroparticle to fundamental physics. The different devices involved in this research originate a large variety of data, which includes atmospheric data and low-energy events suitable to space-weather studies. 

The Pierre Auger Collaboration upholds the principle that data should be accessible and reused by the widest possible community. This is inspired by the FAIR (Findable, Accessible, Interoperable, and Reusable) principles~\cite{FAIR}. Open access to data requires a complex and continuing effort based on offering support and facilitation. 
This approach includes providing a detailed explanation of the detection techniques and of the events reconstruction and selection methods. Data in the experimental proprietary format are translated in portable and flexible files as JSON (JavaScript Object Notation) and CSV (comma-separated values). Finally tutorial and analysis codes are provided as Jupyter Notebooks in Python for easy manipulation.
The implementation of the data release procedure required the approval, by the Collaboration Board, of the \href{https://opendata.auger.org/AugerOpenDataPolicy.pdf}{Data open-access policy of the Pierre Auger Observatory}, and the creation of a dedicated task, under the responsibility of the Project Management, coordinating the continuous effort for releasing data in synergy with the involved physics tasks.

\begin{figure}[t!]%
\vspace{0.2 cm}
\centering
\includegraphics[width=0.85\textwidth]{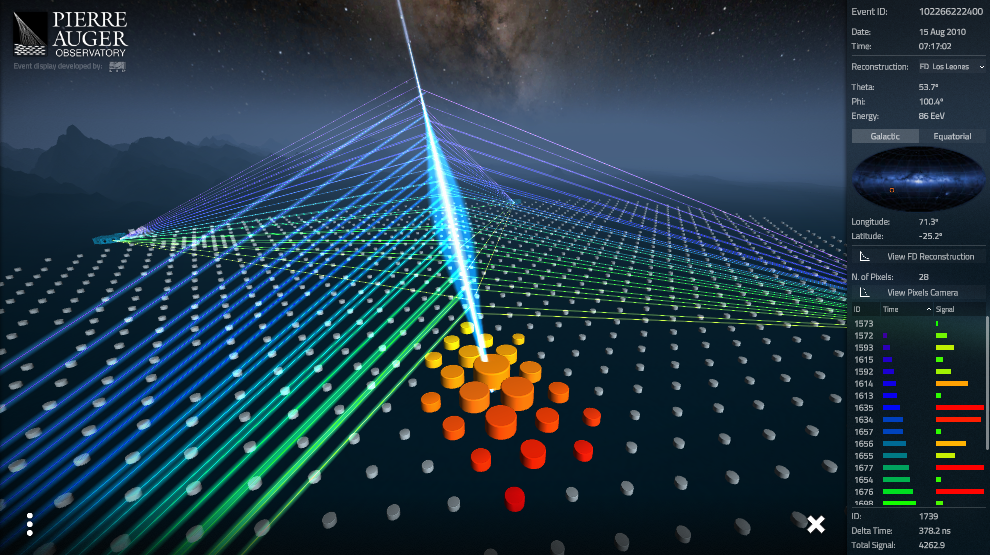}
\caption{Immersive view of the highest energy hybrid event in the Auger Open Data UHECR catalog, PAO100815, with an energy of 82 EeV and zenith of 54 degrees. It triggered 22 surface detector stations and fluorescence telescopes in all 4 fluorescence detector sites. \label{fig:EVhyb}}
\end{figure}

\section {The Auger Open Data Portal}

Following the approved policy, the Open Data Portal~\cite{odp} was set up in February 2021, towards the end of the first phase of operation of the Observatory. Since then, updates were implemented by the Open Data task in synergy with the detector performance and physics analysis groups, and the diversity and quantity of data were enlarged. 

The portal contains 10\% of the cosmic-ray data used for the analyses presented in 2019 at the 36$^\mathrm{th}$ International Cosmic Ray Conference in Madison, Wisconsin, US, and in recent publications, comprising events from both the surface and the fluorescence detectors. An immersive view of the most energetic hybrid 
event, recorded by the surface and fluorescence detectors simultaneously, is provided in Fig.~\ref{fig:EVhyb}. By means of the provided visualisation and analysis tools, the user can select and browse data, and understand details behind the published physics results and reproduce them. The motivations and challenges of the Auger Open Data along with the implementation of the portal and the evolution of its content have been described in detail in~\cite{odEPJC} and presented at previous conferences~\cite{odCONF}. 

In March 2024, data from the low energy extension of the surface detector and the high elevation fluorescence telescopes were added. At present a total of more than 80\,000 showers ranging from an energy of 0.1 EeV up to the highest detected events, and with an angular range representative of the full exposure of the Pierre Auger Observatory, are available for inspection and download. Moreover a browser for the 100 highest-energy events~\cite{catalog} recorded by the surface detector, along with the nine highest-energy hybrid events used for their calibration, has been also implemented. 

All datasets have been produced with the most up-to-date reconstruction software at the time of their release under the (CC BY-SA 4.0) International License, and are identified by a Digital Object Identifier (DOI)~\cite{zenodoauger}. The user is requested to cite this general link, always pointing to the current version, or the link to the specific version of the used data in any applications or publications. 

\begin{figure}[t!]%
\vspace{0.2 cm}
\centering
\includegraphics[width=0.95\textwidth]{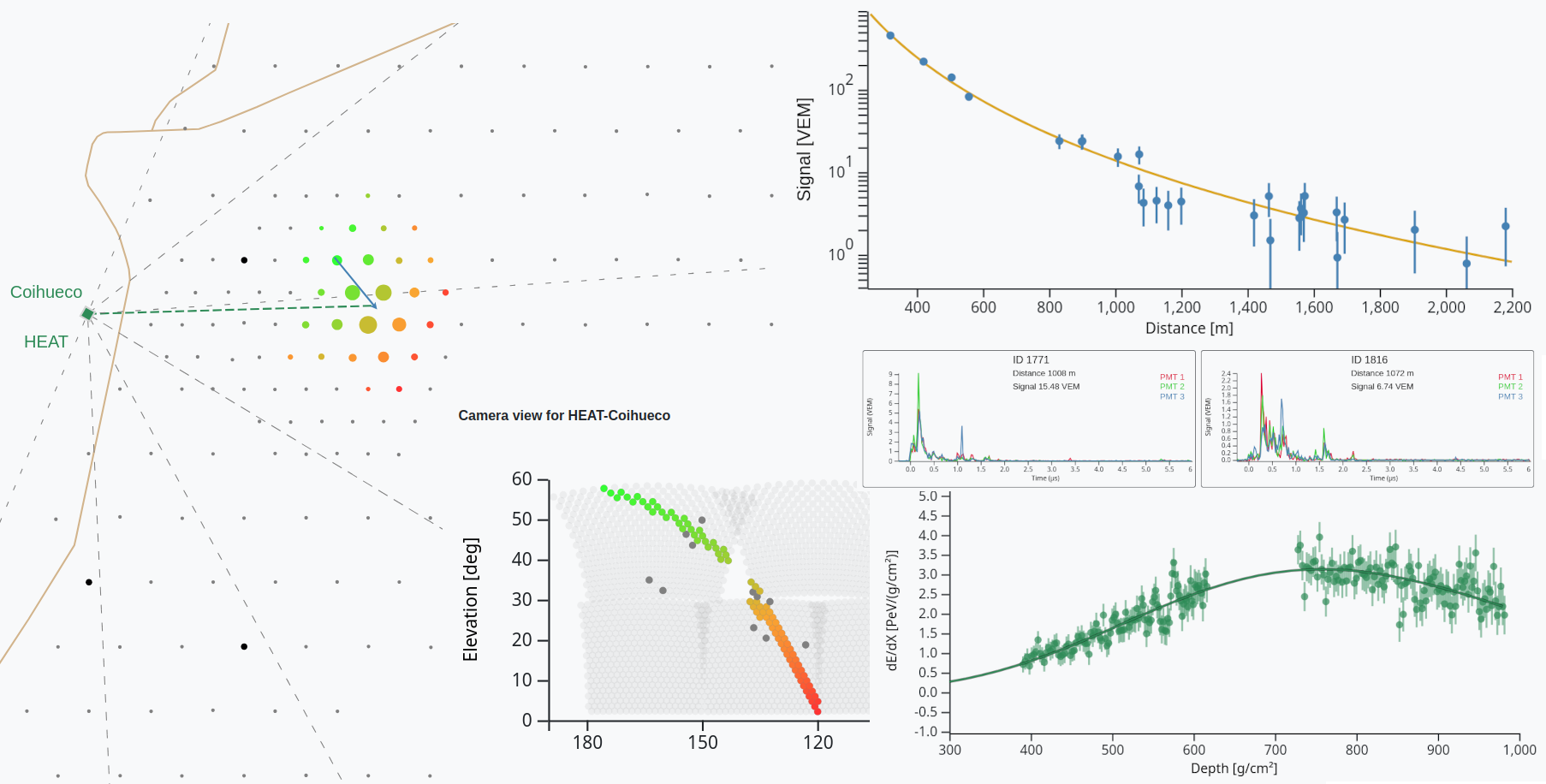}
\caption{Left panels: shower footprint at the ground and camera view of the fluorescence detector for an event of the low energy sample. Right panels: reconstruction of the event by the surface (top) and fluorescence (bottom) detectors. \label{fig:EVlow}}
\vspace{0.5 cm}
\end{figure}

\subsection{The low energy sample}
\label{lowen}

Between 2008 and 2011, the low energy extension of the Pierre Auger Observatory surface detector, dedicated to the measure of cosmic rays near the feature known as the "second knee", was installed by nesting additional water Cherenkov detectors (WCDs) within the surface detector at a mutual distance of 750 meters, forming a denser array, SD750. Moreover, three high elevation air-fluorescence telescopes, HEAT, overlooking the dense array were added. 

The low energy cosmic ray data sample contains 10\% of the data collected by the SD750 with a reconstructed angle below 40 degrees and energy threshold of 0.1 EeV. Around 54\,500 selected events belonging to the dataset published in~\cite{sd750} are available for visualization, manipulation and download.
An exemplary event recorded by the SD750 array and simultaneously seen by the HEAT fluorescence telescopes is displayed in Fig.~\ref{fig:EVlow}. The shower footprint at the ground, along with recorded WCD traces, and reconstruction plots of the surface and fluorescence detectors are shown.

\subsection{The catalog of the highest energy events}
\label{catalog}

The catalog of the 100 highest-energy cosmic-ray events detected during the first phase of the Observatory's operation, between 2004 and 2022, has been published in~\cite{catalog}. The events have a reconstructed energy between 76 EeV and 166 EeV and have been used to study the arrival directions of cosmic rays with energy above 32 EeV~\cite{arrdir}. Full details of the top 100 events are available for inspection and download, along with other nine highest-energy hybrid events used for the energy calibration, in the UHECR catalog section of the Portal at \href{https://opendata.auger.org/catalog/}{https://opendata.auger.org/catalog/}. They are identified with an event name, PAOyymmdd, indicating the year, month, and day of their detection.

Reconstruction parameters, such as Coordinated Universal Time (UTC), energy, zenith and azimuth angles, declination and right ascension, and multiplicity of triggered stations, are available in the event summary table.  Additional features, such as the footprint at the ground and its projection on the shower plane, the lateral distribution of the shower particles, and the time delays of the signals with respect to a plane shower front, can be displayed. For hybrid events, the quantities measured with the fluorescence detector, such as energy and depth of shower maximum are also available. The event files in JSON format also contain the calibrated traces for each photomultiplier tube in the triggered stations. 

\begin{figure}[t!]%
\vspace{0.2 cm}
\centering
\includegraphics[width=0.95\textwidth]{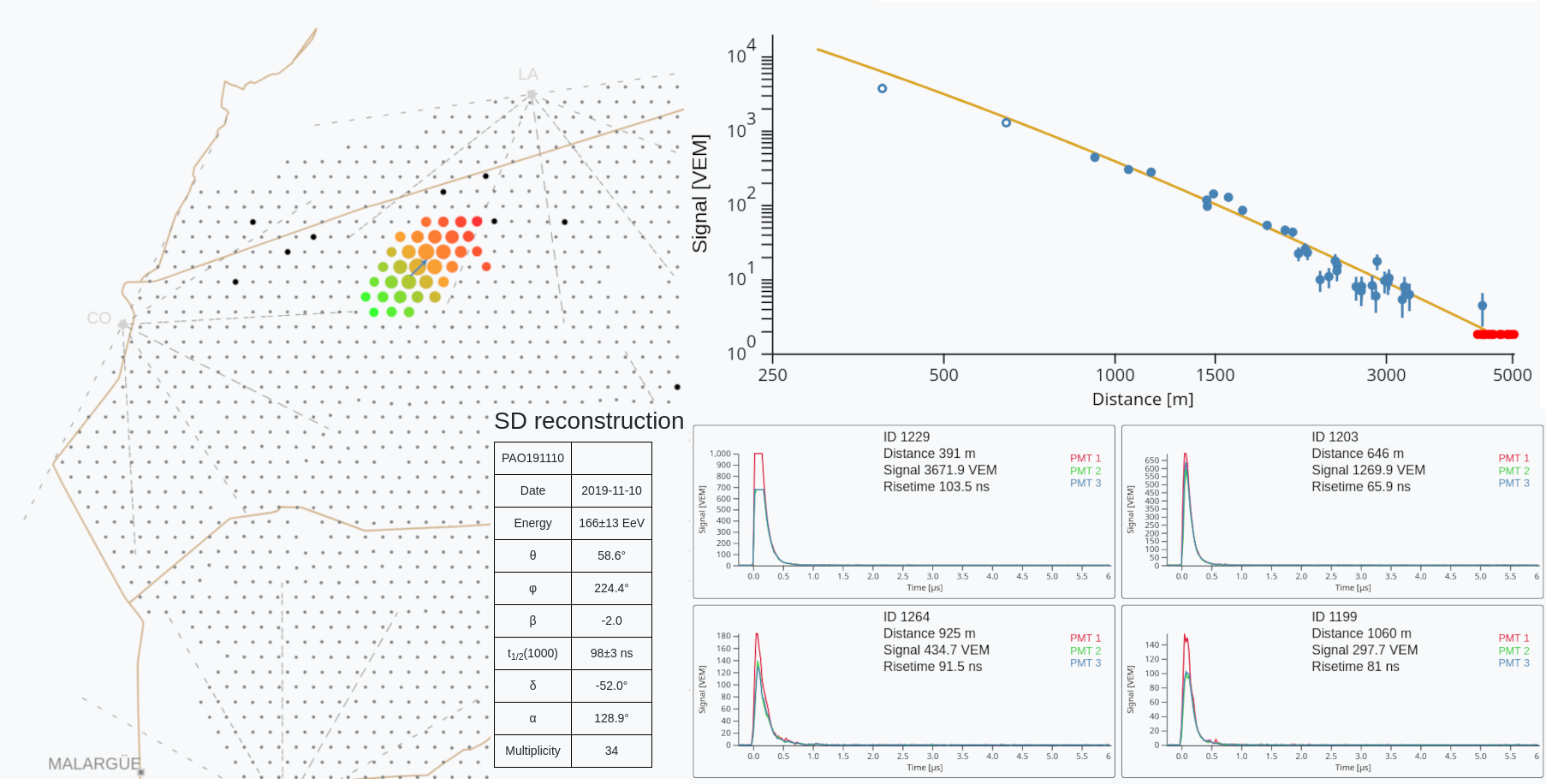}
\caption{The highest energy vertical event, PAO191110: footprint on the ground with event details (left), lateral distribution of the recorded signal as a function of distance to the shower core (top right panel), traces recorded in WCD stations at different distances (bottom right panel).}\label{fig:EVtop}
\vspace{0.5 cm}
\end{figure}

\emph{Vertical events:} the reconstruction of vertical events (with zenith angle below 60 degrees) is described in~\cite{sdrec}. Some properties of the most energetic air shower registered with the surface array, PAO191110, are shown in Fig.~\ref{fig:EVtop}. The primary energy is (166 $\pm$ 13) EeV with the shower impacting the surface array at a zenith angle of 58 degrees. The event footprint on the ground spans an area of approximately (13 $\times$ 6) km$^2$, with 34 WCDs triggered.

\begin{figure}[t!]%
\vspace{0.2 cm}
\centering
\includegraphics[width=0.95\textwidth]{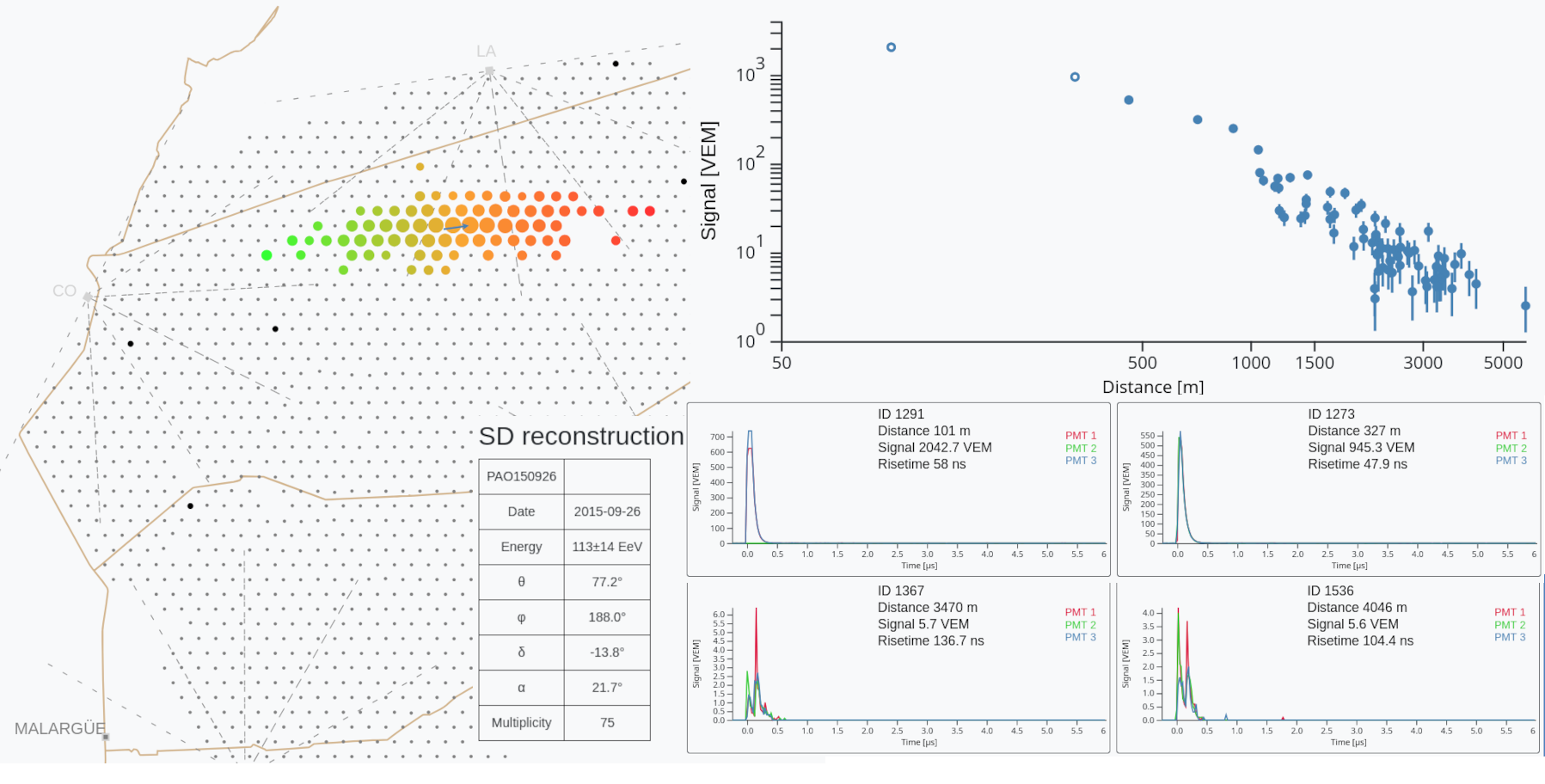}
\caption{The highest energy inclined event, PAO150926: footprint on the ground with event details (left), lateral distribution of the recorded signal as a function of distance to the shower core (top right panel), traces recorded in WCD stations at different distances (bottom right panel).}\label{fig:EVincl}
\vspace{0.5 cm}
\end{figure}

\emph{Inclined events:} the analysis of inclined events (events with zenith angles larger than 60 degrees) is important for extending the sky coverage of the Observatory. The addition of this sample enhances the exposure of the Observatory by 30\%. The reconstruction of events above 60 degrees has been described in~\cite{SDincl}. Details of the highest energy inclined event, PAO150926, are shown in Fig.~\ref{fig:EVincl}. It has a zenith angle of 77 degrees and energy of (113 $\pm$ 14) EeV. The shower footprint is about (35 $\times$ 6) km$^2$ and triggered 75 WCDs in an elongated pattern on the ground. 

\begin{figure}[t!]%
\vspace{0.2 cm}
\centering
\includegraphics[width=0.95\textwidth]{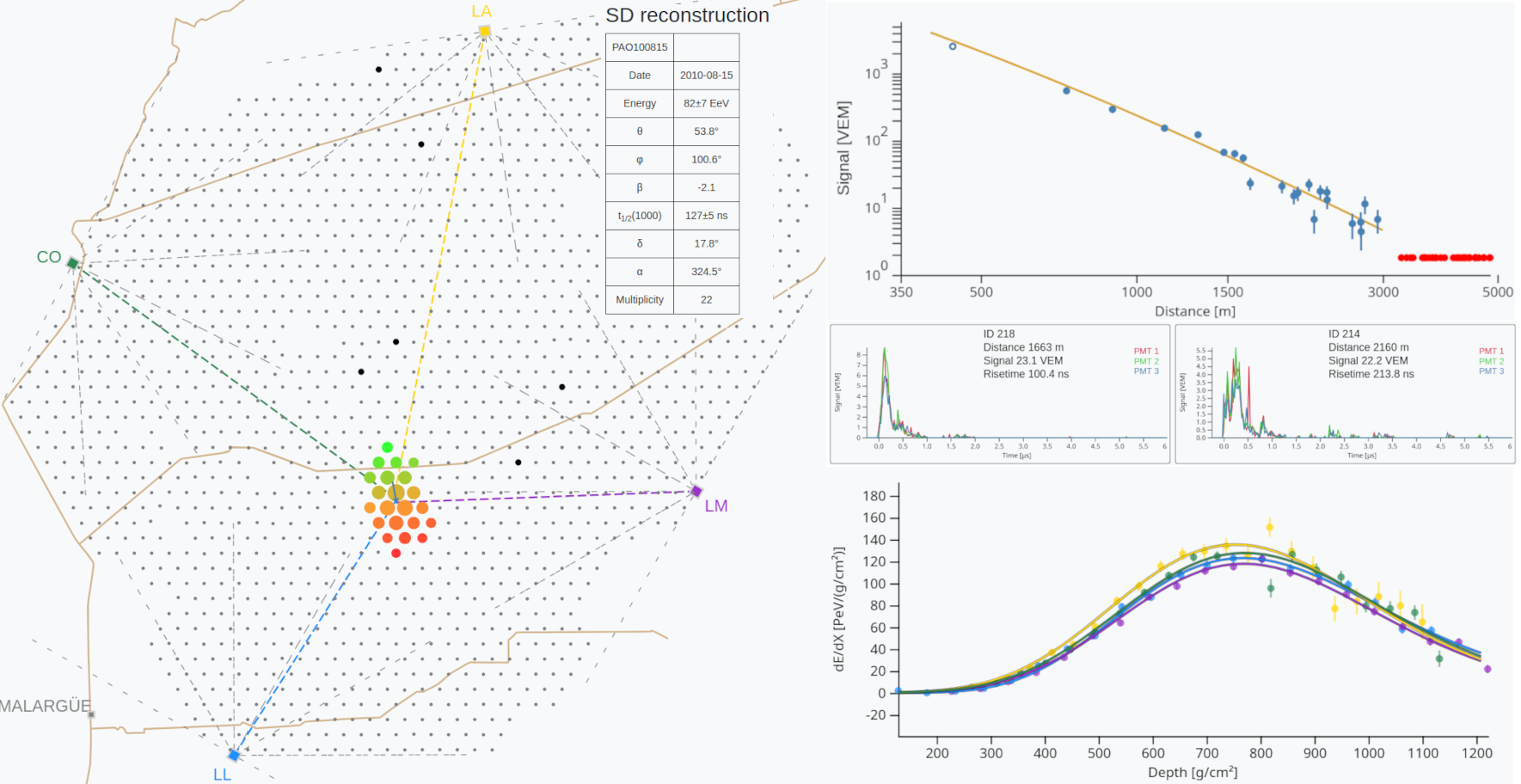}
\caption{The highest-energy multi-eye hybrid event in the UHECR catalog~\cite{catalog}, PAO100815: event footprint projected onto the shower plane (left panel); lateral distribution of the signals as a function of the distance from the shower axis and exemplary WCD traces (top right panel); reconstructed energy deposited in the atmosphere as recorded by the fluorescence detectors (bottom right panel).}\label{fig:EVhybrec}
\vspace{0.5 cm}
\end{figure}

\emph{Hybrid events}: the details of the 10 most energetic hybrid events used for the energy calibration of the full data sample have also been published on the catalog page. The most energetic one is PAO100815, arriving at a zenith angle of 53.8 degrees and with a SD energy estimate of (82 $\pm$ 7) EeV, consistent with the estimate from the FD of (85 $\pm$ 4) EeV. The event had a footprint on the ground of about (7.5 $\times$ 6) km$^2$ and triggered 22 stations. The 3D view of the highest-energy hybrid event in the UHECR catalog is shown in Fig.~\ref{fig:EVhyb}. More details of the reconstruction of the surface and fluorescence detectors are shown in Fig.~\ref{fig:EVhybrec}. 

The UHECR catalog once more demonstrates the quality of the data that lie behind measurements that have been reported by the Pierre Auger Collaboration in recent publications. The full release of the 100 highest-energy events is in line with the Collaboration's commitment to sharing its data and results with the scientific community and to promote the exchange of knowledge between experiments.

\begin{figure*}[hbt!]%
\centering
\includegraphics[width=0.48\textwidth]{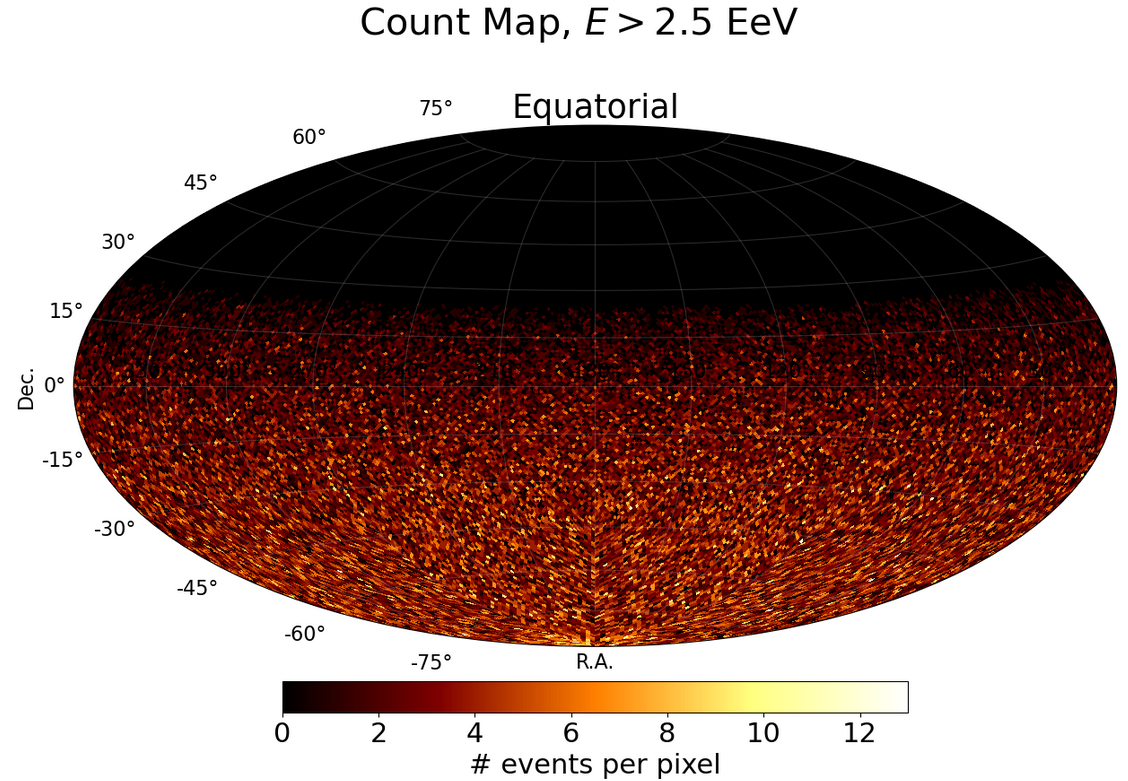}
\includegraphics[width=0.50\textwidth]{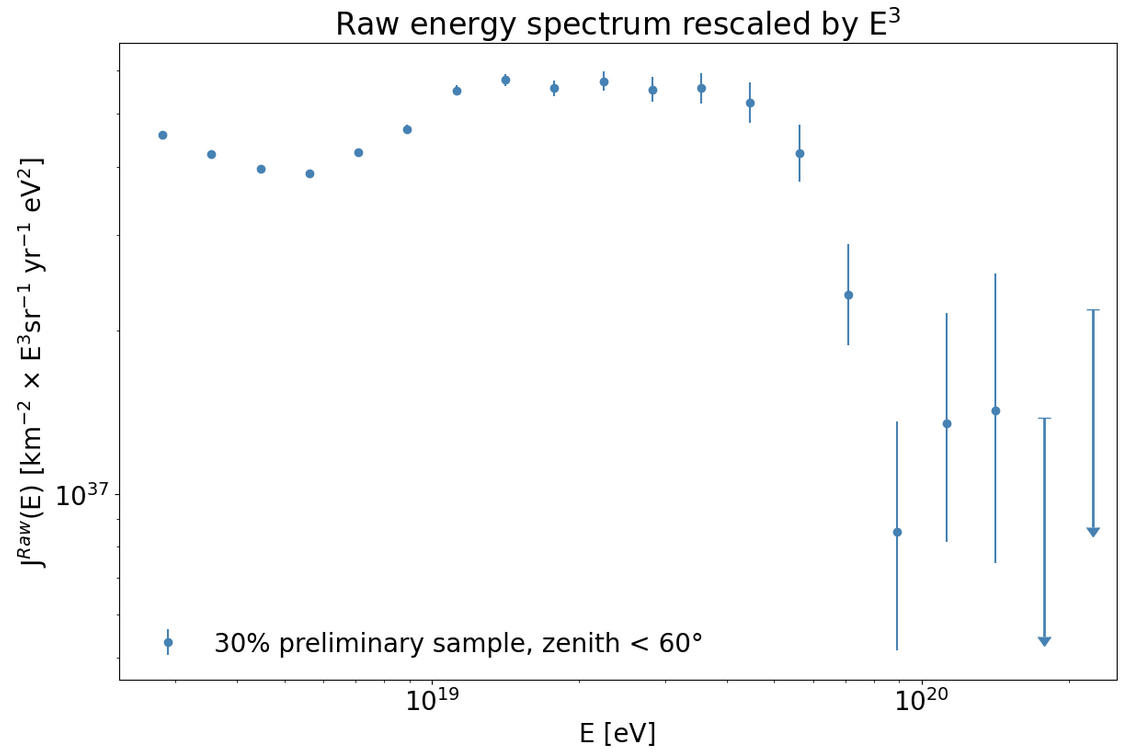}
\caption{Exemplary output of the Python notebooks provided on the Portal applied to the preliminary 30\% sample of selected events with energy above 2.5 EeV and zenith below 60 degrees. Left panel: count map per pixel in equatorial coordinates, mean bin distance $\sim 1$°. Right panel, raw energy spectrum scaled by E$^{3}$.\label{figNB}}
\label{ana30}
\end{figure*}

\section{Expanding the Open Data program: increase of the public CR dataset fraction}

In June 2023, the Pierre Auger Collaboration Board approved the increase of the fraction of released cosmic-ray data to 30\% of the events, collected with the main surface detector array in the period Jan 2004 to Dec 2022. The new release is planned for late 2025. Data is selected by the same criteria applied for the vertical spectrum analysis and will include 30\% of hybrid events used for the energy calibration~\cite{spectrum, spec25}. 

This amounts to an unprecedented exposure of more than 24\,000 km$^2$ sr yr, that will be available for the scientific community. Data will be shared in the same format as the previous releases. 

An example output produced by the Python notebooks available on the Portal, applied to the preliminarily selected data sample of the upcoming 30\% release of selected events with energy above 2.5 EeV and zenith below 60 degrees is shown in Fig.~\ref{ana30}. 
In the left panel the count map of the events per pixel in equatorial coordinates is calculated via the Healpix library with nside = 64, resulting in $\sim$ 49\,000 pixels with mean bin distance of $\sim 1$° and area of 2.6 $\cdot 10^{-4}$ sr. In the right panel, the raw energy spectrum for the selected events, scaled by $E^{3}$, is displayed. The features of the energy spectrum, ankle, suppression and {\em instep}~\cite{spfeat}, are clearly visible. 

\section {Conclusions: open data impact and perspectives}

Open data provides a foundation for sharing knowledge within the scientific community, facilitating joint analyses between different experiments and multi-messenger campaigns that bring together the astroparticle and astronomy communities. 

A few scientific papers using the Auger Open Data have already appeared in journals or on arXiv. Furthermore, open data provide material for developing various activities dedicated to the general public, as well as for high school and university students, which focus on learning physics through hands-on programming and data analysis. The data have also been exploited in outreach events such as the International Cosmic Day, organized by DESY, the IPPOG International Masterclasses program involving more than ten thousand 15- to 19-year-old students from 60 countries. For a comprehensive review of the outreach activities at the Pierre Auger Observatory, please refer to~\cite{outrCONF}.

The use of the released open data is tracked directly via the Zenodo link (\href{https://zenodo.org}{https://zenodo.org}) and with Matomo tools (\href{https://matomo.org}{https://matomo.org}). Since 2021, the portal has received more than 70\,000 visits worldwide, with over 8\,000 lasting longer than one minute. Meanwhile, downloads of data samples number more than 4\,300.

After the endorsement by the Collaboration Board, the next release will disclose 30\% of {\em Phase I} data collected with the surface detector and selected for the vertical spectrum analysis, along with the hybrid events used for the energy calibration. 
This amounts to an exposure of more than 24\,000 km$^2$ sr yr. The Collaboration members are convinced that this will enormously boost the use of the Observatory data by the scientific community. In particular, this data will provide an invaluable source for deeper and wider scientific efforts, thanks to the participation of the Pierre Auger Observatory in world-wide alert systems, Virtual Observatories and coordinated research centres as the consortium ACME~\cite{ACME} recently funded by the European community. 

The Pierre Auger Observatory has recently been upgraded with additional detectors, such as surface detector scintillators, underground muon detectors, radio antennas, and with faster electronics added to each surface detector station~\cite{upgrade}. Future data from the upgraded Observatory, {\em Phase II} data, can be easily integrated into the implemented open data framework towards their gradual release.

%% Full authors list (ONLY FOR COLLABORATIONS)
%\clearpage
%\section*{Full Authors List: \Coll\ Collaboration}
%
%\noindent \textbf{Note comment afterwards:} Collaborations have the possibility to provide an authors list in xml format which will be used while generating the DOI entries making the full authors list searchable in databases like Inspire HEP. \\
%
%\scriptsize
%\noindent
%first.author$^1$, 
%second.author$^2$, 
%third.author$^3$ % .... more names
%and 
%last.author$^{n}$ \\
%
%\noindent
%$^1$first.affiliation.
%$^2$second.affiliation. % .... more affiliation
%$^{m}$last.affiliation.

%\input{latex_authorlist_authors}
%\input{latex_authorlist_institutions}
%\input{acknowledgments}

%----------------------------------
\newpage

\par\noindent
\textbf{The Pierre Auger Collaboration}\\

\begin{wrapfigure}[8]{l}{0.12\linewidth}
\vspace{-2.9ex}
\includegraphics[width=0.98\linewidth]{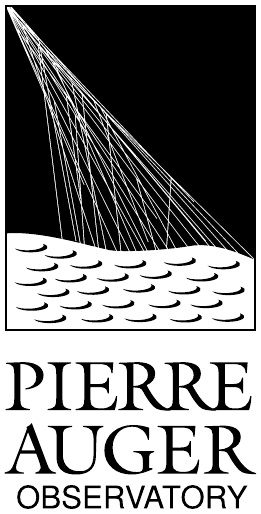}
\end{wrapfigure}
\begin{sloppypar}\noindent
% created on 2025-06-06
A.~Abdul Halim$^{13}$,
P.~Abreu$^{70}$,
M.~Aglietta$^{53,51}$,
I.~Allekotte$^{1}$,
K.~Almeida Cheminant$^{78,77}$,
A.~Almela$^{7,12}$,
R.~Aloisio$^{44,45}$,
J.~Alvarez-Mu\~niz$^{76}$,
A.~Ambrosone$^{44}$,
J.~Ammerman Yebra$^{76}$
% ...

\end{sloppypar}

% created on 2025-06-06
% needs \usepackage{enumitem}
\begin{description}[labelsep=0.2em,align=right,labelwidth=0.7em,labelindent=0em,leftmargin=2em,noitemsep,before={\renewcommand\makelabel[1]{##1 }}]

\item[$^{1}$] Centro At\'omico Bariloche and Instituto Balseiro (CNEA-UNCuyo-CONICET), San Carlos de Bariloche, Argentina
\item[$^{2}$] Departamento de F\'\i{}sica and Departamento de Ciencias de la Atm\'osfera y los Oc\'eanos, FCEyN, Universidad de Buenos Aires and CONICET, Buenos Aires, Argentina
\item[$^{3}$] IFLP, Universidad Nacional de La Plata and CONICET, La Plata, Argentina
\item[$^{4}$] Instituto de Astronom\'\i{}a y F\'\i{}sica del Espacio (IAFE, CONICET-UBA), Buenos Aires, Argentina
\item[$^{5}$] Instituto de F\'\i{}sica de Rosario (IFIR) -- CONICET/U.N.R.\ and Facultad de Ciencias Bioqu\'\i{}micas y Farmac\'euticas U.N.R., Rosario, Argentina
\item[$^{6}$] Instituto de Tecnolog\'\i{}as en Detecci\'on y Astropart\'\i{}culas (CNEA, CONICET, UNSAM), and Universidad Tecnol\'ogica Nacional -- Facultad Regional Mendoza (CONICET/CNEA), Mendoza, Argentina
\item[$^{7}$] Instituto de Tecnolog\'\i{}as en Detecci\'on y Astropart\'\i{}culas (CNEA, CONICET, UNSAM), Buenos Aires, Argentina
\item[$^{8}$] International Center of Advanced Studies and Instituto de Ciencias F\'\i{}sicas, ECyT-UNSAM and CONICET, Campus Miguelete -- San Mart\'\i{}n, Buenos Aires, Argentina
\item[$^{9}$] Laboratorio Atm\'osfera -- Departamento de Investigaciones en L\'aseres y sus Aplicaciones -- UNIDEF (CITEDEF-CONICET), Argentina
\item[$^{10}$] Observatorio Pierre Auger, Malarg\"ue, Argentina
\item[$^{11}$] Observatorio Pierre Auger and Comisi\'on Nacional de Energ\'\i{}a At\'omica, Malarg\"ue, Argentina
\item[$^{12}$] Universidad Tecnol\'ogica Nacional -- Facultad Regional Buenos Aires, Buenos Aires, Argentina
\item[$^{13}$] University of Adelaide, Adelaide, S.A., Australia
\item[$^{14}$] Universit\'e Libre de Bruxelles (ULB), Brussels, Belgium
\item[$^{15}$] Vrije Universiteit Brussels, Brussels, Belgium
\item[$^{16}$] Centro Brasileiro de Pesquisas Fisicas, Rio de Janeiro, RJ, Brazil
\item[$^{17}$] Centro Federal de Educa\c{c}\~ao Tecnol\'ogica Celso Suckow da Fonseca, Petropolis, Brazil
\item[$^{18}$] Instituto Federal de Educa\c{c}\~ao, Ci\^encia e Tecnologia do Rio de Janeiro (IFRJ), Brazil
\item[$^{19}$] Universidade de S\~ao Paulo, Escola de Engenharia de Lorena, Lorena, SP, Brazil
\item[$^{20}$] Universidade de S\~ao Paulo, Instituto de F\'\i{}sica de S\~ao Carlos, S\~ao Carlos, SP, Brazil
\item[$^{21}$] Universidade de S\~ao Paulo, Instituto de F\'\i{}sica, S\~ao Paulo, SP, Brazil
\item[$^{22}$] Universidade Estadual de Campinas (UNICAMP), IFGW, Campinas, SP, Brazil
\item[$^{23}$] Universidade Estadual de Feira de Santana, Feira de Santana, Brazil
\item[$^{24}$] Universidade Federal de Campina Grande, Centro de Ciencias e Tecnologia, Campina Grande, Brazil
\item[$^{25}$] Universidade Federal do ABC, Santo Andr\'e, SP, Brazil
\item[$^{26}$] Universidade Federal do Paran\'a, Setor Palotina, Palotina, Brazil
\item[$^{27}$] Universidade Federal do Rio de Janeiro, Instituto de F\'\i{}sica, Rio de Janeiro, RJ, Brazil
\item[$^{28}$] Universidad de Medell\'\i{}n, Medell\'\i{}n, Colombia
\item[$^{29}$] Universidad Industrial de Santander, Bucaramanga, Colombia
\item[$^{30}$] Charles University, Faculty of Mathematics and Physics, Institute of Particle and Nuclear Physics, Prague, Czech Republic
\item[$^{31}$] Institute of Physics of the Czech Academy of Sciences, Prague, Czech Republic
\item[$^{32}$] Palacky University, Olomouc, Czech Republic
\item[$^{33}$] CNRS/IN2P3, IJCLab, Universit\'e Paris-Saclay, Orsay, France
\item[$^{34}$] Laboratoire de Physique Nucl\'eaire et de Hautes Energies (LPNHE), Sorbonne Universit\'e, Universit\'e de Paris, CNRS-IN2P3, Paris, France
\item[$^{35}$] Univ.\ Grenoble Alpes, CNRS, Grenoble Institute of Engineering Univ.\ Grenoble Alpes, LPSC-IN2P3, 38000 Grenoble, France
\item[$^{36}$] Universit\'e Paris-Saclay, CNRS/IN2P3, IJCLab, Orsay, France
\item[$^{37}$] Bergische Universit\"at Wuppertal, Department of Physics, Wuppertal, Germany
\item[$^{38}$] Karlsruhe Institute of Technology (KIT), Institute for Experimental Particle Physics, Karlsruhe, Germany
\item[$^{39}$] Karlsruhe Institute of Technology (KIT), Institut f\"ur Prozessdatenverarbeitung und Elektronik, Karlsruhe, Germany
\item[$^{40}$] Karlsruhe Institute of Technology (KIT), Institute for Astroparticle Physics, Karlsruhe, Germany
\item[$^{41}$] RWTH Aachen University, III.\ Physikalisches Institut A, Aachen, Germany
\item[$^{42}$] Universit\"at Hamburg, II.\ Institut f\"ur Theoretische Physik, Hamburg, Germany
\item[$^{43}$] Universit\"at Siegen, Department Physik -- Experimentelle Teilchenphysik, Siegen, Germany
\item[$^{44}$] Gran Sasso Science Institute, L'Aquila, Italy
\item[$^{45}$] INFN Laboratori Nazionali del Gran Sasso, Assergi (L'Aquila), Italy
\item[$^{46}$] INFN, Sezione di Catania, Catania, Italy
\item[$^{47}$] INFN, Sezione di Lecce, Lecce, Italy
\item[$^{48}$] INFN, Sezione di Milano, Milano, Italy
\item[$^{49}$] INFN, Sezione di Napoli, Napoli, Italy
\item[$^{50}$] INFN, Sezione di Roma ``Tor Vergata'', Roma, Italy
\item[$^{51}$] INFN, Sezione di Torino, Torino, Italy
\item[$^{52}$] Istituto di Astrofisica Spaziale e Fisica Cosmica di Palermo (INAF), Palermo, Italy
\item[$^{53}$] Osservatorio Astrofisico di Torino (INAF), Torino, Italy
\item[$^{54}$] Politecnico di Milano, Dipartimento di Scienze e Tecnologie Aerospaziali , Milano, Italy
\item[$^{55}$] Universit\`a del Salento, Dipartimento di Matematica e Fisica ``E.\ De Giorgi'', Lecce, Italy
\item[$^{56}$] Universit\`a dell'Aquila, Dipartimento di Scienze Fisiche e Chimiche, L'Aquila, Italy
\item[$^{57}$] Universit\`a di Catania, Dipartimento di Fisica e Astronomia ``Ettore Majorana``, Catania, Italy
\item[$^{58}$] Universit\`a di Milano, Dipartimento di Fisica, Milano, Italy
\item[$^{59}$] Universit\`a di Napoli ``Federico II'', Dipartimento di Fisica ``Ettore Pancini'', Napoli, Italy
\item[$^{60}$] Universit\`a di Palermo, Dipartimento di Fisica e Chimica ''E.\ Segr\`e'', Palermo, Italy
\item[$^{61}$] Universit\`a di Roma ``Tor Vergata'', Dipartimento di Fisica, Roma, Italy
\item[$^{62}$] Universit\`a Torino, Dipartimento di Fisica, Torino, Italy
\item[$^{63}$] Benem\'erita Universidad Aut\'onoma de Puebla, Puebla, M\'exico
\item[$^{64}$] Unidad Profesional Interdisciplinaria en Ingenier\'\i{}a y Tecnolog\'\i{}as Avanzadas del Instituto Polit\'ecnico Nacional (UPIITA-IPN), M\'exico, D.F., M\'exico
\item[$^{65}$] Universidad Aut\'onoma de Chiapas, Tuxtla Guti\'errez, Chiapas, M\'exico
\item[$^{66}$] Universidad Michoacana de San Nicol\'as de Hidalgo, Morelia, Michoac\'an, M\'exico
\item[$^{67}$] Universidad Nacional Aut\'onoma de M\'exico, M\'exico, D.F., M\'exico
\item[$^{68}$] Institute of Nuclear Physics PAN, Krakow, Poland
\item[$^{69}$] University of \L{}\'od\'z, Faculty of High-Energy Astrophysics,\L{}\'od\'z, Poland
\item[$^{70}$] Laborat\'orio de Instrumenta\c{c}\~ao e F\'\i{}sica Experimental de Part\'\i{}culas -- LIP and Instituto Superior T\'ecnico -- IST, Universidade de Lisboa -- UL, Lisboa, Portugal
\item[$^{71}$] ``Horia Hulubei'' National Institute for Physics and Nuclear Engineering, Bucharest-Magurele, Romania
\item[$^{72}$] Institute of Space Science, Bucharest-Magurele, Romania
\item[$^{73}$] Center for Astrophysics and Cosmology (CAC), University of Nova Gorica, Nova Gorica, Slovenia
\item[$^{74}$] Experimental Particle Physics Department, J.\ Stefan Institute, Ljubljana, Slovenia
\item[$^{75}$] Universidad de Granada and C.A.F.P.E., Granada, Spain
\item[$^{76}$] Instituto Galego de F\'\i{}sica de Altas Enerx\'\i{}as (IGFAE), Universidade de Santiago de Compostela, Santiago de Compostela, Spain
\item[$^{77}$] IMAPP, Radboud University Nijmegen, Nijmegen, The Netherlands
\item[$^{78}$] Nationaal Instituut voor Kernfysica en Hoge Energie Fysica (NIKHEF), Science Park, Amsterdam, The Netherlands
\item[$^{79}$] Stichting Astronomisch Onderzoek in Nederland (ASTRON), Dwingeloo, The Netherlands
\item[$^{80}$] Universiteit van Amsterdam, Faculty of Science, Amsterdam, The Netherlands
\item[$^{81}$] Case Western Reserve University, Cleveland, OH, USA
\item[$^{82}$] Colorado School of Mines, Golden, CO, USA
\item[$^{83}$] Department of Physics and Astronomy, Lehman College, City University of New York, Bronx, NY, USA
\item[$^{84}$] Michigan Technological University, Houghton, MI, USA
\item[$^{85}$] New York University, New York, NY, USA
\item[$^{86}$] University of Chicago, Enrico Fermi Institute, Chicago, IL, USA
\item[$^{87}$] University of Delaware, Department of Physics and Astronomy, Bartol Research Institute, Newark, DE, USA
\item[] -----
\item[$^{a}$] Max-Planck-Institut f\"ur Radioastronomie, Bonn, Germany
\item[$^{b}$] also at Kapteyn Institute, University of Groningen, Groningen, The Netherlands
\item[$^{c}$] School of Physics and Astronomy, University of Leeds, Leeds, United Kingdom
\item[$^{d}$] Fermi National Accelerator Laboratory, Fermilab, Batavia, IL, USA
\item[$^{e}$] Pennsylvania State University, University Park, PA, USA
\item[$^{f}$] Colorado State University, Fort Collins, CO, USA
\item[$^{g}$] Louisiana State University, Baton Rouge, LA, USA
\item[$^{h}$] now at Graduate School of Science, Osaka Metropolitan University, Osaka, Japan
\item[$^{i}$] Institut universitaire de France (IUF), France
\item[$^{j}$] now at Technische Universit\"at Dortmund and Ruhr-Universit\"at Bochum, Dortmund and Bochum, Germany
\end{description}

% created on 2025-06-06
\section*{Acknowledgments}

\begin{sloppypar}
The successful installation, commissioning, and operation of the Pierre
Auger Observatory would not have been possible without the strong
commitment and effort from the technical and administrative staff in
Malarg\"ue. We are very grateful to the following agencies and
organizations for financial support:
\end{sloppypar}

\begin{sloppypar}
Argentina -- Comisi\'on Nacional de Energ\'\i{}a At\'omica; Agencia Nacional de
Promoci\'on Cient\'\i{}fica y Tecnol\'ogica (ANPCyT); Consejo Nacional de
Investigaciones Cient\'\i{}ficas y T\'ecnicas (CONICET); Gobierno de la
Provincia de Mendoza; Municipalidad de Malarg\"ue; NDM Holdings and Valle
Las Le\~nas; in gratitude for their continuing cooperation over land
access; Australia -- the Australian Research Council; Belgium -- Fonds
de la Recherche Scientifique (FNRS); Research Foundation Flanders (FWO),
Marie Curie Action of the European Union Grant No.~101107047; Brazil --
Conselho Nacional de Desenvolvimento Cient\'\i{}fico e Tecnol\'ogico (CNPq);
Financiadora de Estudos e Projetos (FINEP); Funda\c{c}\~ao de Amparo \`a
Pesquisa do Estado de Rio de Janeiro (FAPERJ); S\~ao Paulo Research
Foundation (FAPESP) Grants No.~2019/10151-2, No.~2010/07359-6 and
No.~1999/05404-3; Minist\'erio da Ci\^encia, Tecnologia, Inova\c{c}\~oes e
Comunica\c{c}\~oes (MCTIC); Czech Republic -- GACR 24-13049S, CAS LQ100102401,
MEYS LM2023032, CZ.02.1.01/0.0/0.0/16{\textunderscore}013/0001402,
CZ.02.1.01/0.0/0.0/18{\textunderscore}046/0016010 and
CZ.02.1.01/0.0/0.0/17{\textunderscore}049/0008422 and CZ.02.01.01/00/22{\textunderscore}008/0004632;
France -- Centre de Calcul IN2P3/CNRS; Centre National de la Recherche
Scientifique (CNRS); Conseil R\'egional Ile-de-France; D\'epartement
Physique Nucl\'eaire et Corpusculaire (PNC-IN2P3/CNRS); D\'epartement
Sciences de l'Univers (SDU-INSU/CNRS); Institut Lagrange de Paris (ILP)
Grant No.~LABEX ANR-10-LABX-63 within the Investissements d'Avenir
Programme Grant No.~ANR-11-IDEX-0004-02; Germany -- Bundesministerium
f\"ur Bildung und Forschung (BMBF); Deutsche Forschungsgemeinschaft (DFG);
Finanzministerium Baden-W\"urttemberg; Helmholtz Alliance for
Astroparticle Physics (HAP); Helmholtz-Gemeinschaft Deutscher
Forschungszentren (HGF); Ministerium f\"ur Kultur und Wissenschaft des
Landes Nordrhein-Westfalen; Ministerium f\"ur Wissenschaft, Forschung und
Kunst des Landes Baden-W\"urttemberg; Italy -- Istituto Nazionale di
Fisica Nucleare (INFN); Istituto Nazionale di Astrofisica (INAF);
Ministero dell'Universit\`a e della Ricerca (MUR); CETEMPS Center of
Excellence; Ministero degli Affari Esteri (MAE), ICSC Centro Nazionale
di Ricerca in High Performance Computing, Big Data and Quantum
Computing, funded by European Union NextGenerationEU, reference code
CN{\textunderscore}00000013; M\'exico -- Consejo Nacional de Ciencia y Tecnolog\'\i{}a
(CONACYT) No.~167733; Universidad Nacional Aut\'onoma de M\'exico (UNAM);
PAPIIT DGAPA-UNAM; The Netherlands -- Ministry of Education, Culture and
Science; Netherlands Organisation for Scientific Research (NWO); Dutch
national e-infrastructure with the support of SURF Cooperative; Poland
-- Ministry of Education and Science, grants No.~DIR/WK/2018/11 and
2022/WK/12; National Science Centre, grants No.~2016/22/M/ST9/00198,
2016/23/B/ST9/01635, 2020/39/B/ST9/01398, and 2022/45/B/ST9/02163;
Portugal -- Portuguese national funds and FEDER funds within Programa
Operacional Factores de Competitividade through Funda\c{c}\~ao para a Ci\^encia
e a Tecnologia (COMPETE); Romania -- Ministry of Research, Innovation
and Digitization, CNCS-UEFISCDI, contract no.~30N/2023 under Romanian
National Core Program LAPLAS VII, grant no.~PN 23 21 01 02 and project
number PN-III-P1-1.1-TE-2021-0924/TE57/2022, within PNCDI III; Slovenia
-- Slovenian Research Agency, grants P1-0031, P1-0385, I0-0033, N1-0111;
Spain -- Ministerio de Ciencia e Innovaci\'on/Agencia Estatal de
Investigaci\'on (PID2019-105544GB-I00, PID2022-140510NB-I00 and
RYC2019-027017-I), Xunta de Galicia (CIGUS Network of Research Centers,
Consolidaci\'on 2021 GRC GI-2033, ED431C-2021/22 and ED431F-2022/15),
Junta de Andaluc\'\i{}a (SOMM17/6104/UGR and P18-FR-4314), and the European
Union (Marie Sklodowska-Curie 101065027 and ERDF); USA -- Department of
Energy, Contracts No.~DE-AC02-07CH11359, No.~DE-FR02-04ER41300,
No.~DE-FG02-99ER41107 and No.~DE-SC0011689; National Science Foundation,
Grant No.~0450696, and NSF-2013199; The Grainger Foundation; Marie
Curie-IRSES/EPLANET; European Particle Physics Latin American Network;
and UNESCO.
\end{sloppypar}

%----------------------------------


\begin{thebibliography}{99}

\bibitem{paoNIM}{Pierre Auger Collaboration [A. Aab, et al.], %\emph{The Pierre Auger Cosmic Ray Observatory}, 
\href{https://doi.org/10.1016/j.nima.2015.06.058}{Nucl. Instrum. Meth. A {\bf 798} (2015) 172-213}}

\bibitem{FAIR} {Wilkinson, M., et al., %\emph{The FAIR Guiding Principles for scientific data management and stewardship}, 
\href{https://doi.org/10.1038/sdata.2016.18}{Sci Data {\bf 3} (2016) 160018}; and \href{https://openaccess.mpg.de/Berlin-Declaration}{\emph{Berlin Declaration on Open Access to Knowledge in the Sciences and Humanities}}}

%\bibitem{POLICY}{\href{https://opendata.auger.org/AugerOpenDataPolicy.pdf}{\emph{Data open-access policy of the Pierre Auger Observatory}}}

%\bibitem{creative}{Creative Common (CC BY-SA 4.0) Int. License \href{https://creativecommons.org/licenses/by-sa/4.0)}{https://creativecommons.org/licenses/by-sa/4.0)}}

%\bibitem{jupyter}{Jupyter Notebook Interface, \href{https://https://jupyter.org/}{https://https://jupyter.org/}}

\bibitem{odp}{The Pierre Auger Observatory Open Data Portal \href{https://opendata.auger.org/}{https://opendata.auger.org/}}

\bibitem{odEPJC}{Pierre Auger Collaboration [A. Abdul Halim, et al.], %\emph{The Pierre Auger Observatory Open Data}, 
\href{https://doi.org/10.1140/epjc/s10052-024-13560-5}{Eur. Phys. J. C {\bf85} (2025) 70}, \href{https://arxiv.org/abs/2309.16294}{arXiv:2309.16294 [astro-ph]}}

\bibitem{odCONF}{V. Scherini for the Pierre Auger Collaboration, %\emph{The 2021 Open-Data release by the Pierre Auger Collaboration}, 
\href{https://pos.sissa.it/395/1386/}{PoS 37th ICRC (2021) 1386}; P.~L.~Ghia for the Pierre Auger Collaboration, %\emph{Portals to data of the Pierre Auger Observatory}, 
\href{https://pos.sissa.it/444/1616/}{PoS 38th ICRC (2023) 1616}; V.~Scherini for the Pierre Auger Collaboration, \href{https://pos.sissa.it/484/114/}{PoS (UHECR 2024) 114}}


\bibitem{catalog}{Pierre Auger Collaboration [A. Abdul Halim, et al.], %\emph{A Catalog of the Highest-energy Cosmic Rays Recorded during Phase I of Operation of the Pierre Auger Observatory}, 
\href{https://doi.org/10.3847/1538-4365/aca537}{Astrophys. J. Suppl. {\bf 264}, 2 (2023) 50}%, \url{http://arxiv.org/abs/2211.16020}}
}


\bibitem{zenodoauger}{Pierre Auger Observatory Open Data, Zenodo, \href{https://doi.org/10.5281/zenodo.4487612}{https://doi.org/10.5281/zenodo.4487612}}


\bibitem{sd750}{Pierre Auger Collaboration [Abreu, P., et al.]\href{https://doi.org/10.1140/epjc/s10052-021-09700-w}{Eur. Phys. J. C {\bf 81} (2021) 966}}


\bibitem{arrdir}{Pierre Auger Collaboration [P. Abreu, et al.], %\emph{Arrival Directions of Cosmic Rays above 32 EeV from Phase One of the Pierre Auger Observatory}, 
\href{https://doi.org/10.3847/1538-4357/ac7d4e}{ApJ {\bf 935} (2022) 170}}

\bibitem{sdrec}{Pierre Auger Collaboration [Aab, A., et al.]\href{https://doi.org/10.1088/1748-0221/15/10/P10021}{JInst {\bf 15} (2020) P10021}}


\bibitem{SDincl}{Pierre Auger collaboration  \href{https://dx.doi.org/10.1088/1475-7516/2014/08/019}{JCAP {\bf 08} (2014) 019}}


\bibitem{spectrum}{Pierre Auger Collaboration [A. Aab, et al.], %\emph{Measurement of the cosmic-ray energy spectrum above $2.5 \times 10^{18}$ eV using the Pierre Auger Observatory}, 
\href{https://doi.org/10.1103/PhysRevD.102.062005}{Phys. Rev. D  {\bf 102} (2020) 062005}}  % \url{http://arxiv.org/abs/2008.06486}}

\bibitem{spec25}{D.~Ravignani for the Pierre Auger Collaboration, at this conference ICRC2025 \#268}


\bibitem{spfeat}{
Pierre Auger Collaboration [A. Aab, et al.], %\emph{Features of the Energy Spectrum of Cosmic Rays above $2.5 \times 10^{18}$ eV Using the Pierre Auger Observatory}, 
\href{https://doi.org/10.1103/PhysRevLett.125.121106}{Phys. Rev. Lett.  {\bf 125} (2020) 121106}
%, \url{http://arxiv.org/abs/2008.06488}
}

\bibitem{outrCONF}{R.~Sarmento for the Pierre Auger Collaboration, %\emph{International Masterclasses as part of the Pierre Auger Observatory program of Outreach and Education}, 
\href{https://pos.sissa.it/444/1611/}{PoS 38th ICRC (2023) 1611}; B.~Garcia for the Pierre Auger Collaboration, \href{https://pos.sissa.it/484/088/}{PoS (UHECR 2024) 088}; F.~Convenga for the Pierre Auger Collaboration, at this conference ICRC2025 \#797}

\bibitem{ACME}{ACME project, HORIZON-INFRA-2023-SERV-01, Grant Agreement No 101131928. \href{https://doi.org/10.3030/101131928}{https://doi.org/10.3030/101131928}}

\bibitem{upgrade}{Pierre Auger Collaboration [A. Aab, et al.], %\emph{The Pierre Auger Observatory Upgrade - Preliminary Design Report},
\href{https://arxiv.org/abs/1604.03637}{arXiv:1604.03637 [astro-ph]}}


\end{thebibliography}
\end{document}